\documentclass[preprint,final,5p,times,twocolumn]{elsarticle}
\usepackage{amssymb}
\usepackage{graphicx}
\usepackage{multirow}
\usepackage{color}
\usepackage{amsthm}
\usepackage{amsmath}
\usepackage{widetext}
\usepackage{amssymb}
\usepackage{float}
\usepackage[utf8x]{inputenc}
\usepackage[numbers]{natbib}
\usepackage[colorlinks=true,citecolor=blue]{hyperref}
\journal{Physics Letters B}
\begin{document}
\begin{frontmatter}
\title{$w$-mode oscillation of neutron star in a new relativistic hybrid model }
\author{D. Dey$^{1,2*}$}
\ead{*debabrata.d@iopb.res.in}
\author{Jeet Amrit Pattnaik$^{3**}$}
\ead{**jeetamritboudh@gmail.com}
\author{R. N. Panda$^{3}$}
\author{S. K. Patra$^{3}$}
\address[1]{Institute of Physics, Sachivalya Marg, Bhubaneswar-751005, India}
\address[2]{Homi Bhabha National Institute, Training School Complex, 
Anushakti Nagar, Mumbai 400094, India}
\address[3]{Department of Physics, Siksha $'O'$ Anusandhan, Deemed to be University, Bhubaneswar-751030, India}

\begin{abstract}
We investigate how the pulsation frequencies of axial gravitational-wave modes ($w$-modes) in a non-rotating neutron star depend on its composition, particularly when including quarkyonic matter and fermionic dark matter. These modes emerge from the coupling between the star's fluid component and the gravitational field of general relativity, which are highly damped and characterized by complex frequencies with comparable real and imaginary parts. Using a relativistic mean field formalism for the nucleonic component, we modeled the neutron star's interior, while the exterior is analyzed through the complex-coordinate method to determine the $w$-modes. Our study employs a realistic equation of state, based on different physical assumptions and covering a broad area of observational constraints, starting from finite nuclei to nuclear matter with extreme conditions. The numerical findings demonstrate that axial $w$-modes provide valuable insights into the properties of neutron star matter, highlighting their significance in probing the star's internal structure.
\end{abstract}
\end{frontmatter}

\section{Introduction}
\label{sec1}
\noindent
Neutron stars (NSs) serve as extraordinary cosmic laboratories, enabling the study of matter under extreme densities far exceeding nuclear saturation. These ultra-dense objects provide conditions unmatched elsewhere in the universe, making them ideal for testing the behavior of matter at supranuclear densities. Typically, it has a mass $\sim$ 2$M_{\odot}$ and a radius of $\sim$ 12 km \cite{dense_matter_1, dense_matter_2}. Given their immense compactness, a general relativistic framework is essential for accurately modeling their dynamics. Their vibrations are expected to generate continuous gravitational-wave signals, offering a window into the interplay between strong gravity and high-density matter. Among the rich spectrum of non-radial oscillations in NSs, various pulsation modes arise, each governed by distinct restoring forces. The most significant include the f-mode (driven by global fluid oscillations), g-modes (resulting from buoyancy effects), and p-modes (restored by pressure gradients) \cite{fmode2, DDey1}. Additionally, general relativity predicts the existence of $w$-modes—unique space-time oscillations with no Newtonian counterpart \cite{w_mode1,w_mode2, Andersson1998}. These modes, along with the fluid-dominated ones, encode crucial information about the star's internal structure and composition, making them invaluable probes for both astrophysics and gravitational-wave astronomy. \\
The equation of state (EOS) is crucial for understanding the extreme-density matter in NSs. Various theoretical models, both relativistic and non-relativistic, have been developed to determine the EOS. Among these, Relativistic Mean Field Theory (RMF) stands out due to its success in describing nuclear properties across the nuclear chart—from finite and exotic nuclei to NS matter. As a covariant quantum field theory, RMF provides a self-consistent framework for solving the nuclear many-body problem, making it a powerful tool for NS studies \cite{E-RMF1,E-RMF2,E-RMF3,E-RMF4,E-RMF5,E-RMF6,E-RMF7,E-RMF8,E-RMF9}. NSs may harbor exotic matter phases due to their extreme densities. Recent constraints from gravitational-wave event GW170817 \cite{PhysRevLett.119.161101, PhysRevLett.121.091102,PhysRevLett.121.161101, Capano2020} and NICER observations of PSR J0030+0451 \cite{Riley_2019,Miller_2019}  put forward the constraint that a 1.4 $M_{\odot}$ NS has a radius $\leq$ 13.5 km, along with measurements of star masses $\geq$ 2$M_{\odot}$ \cite{Obs1,Obs2} challenged existing dense-matter models \cite{Quarkyonic1}. To explain these observations, McLerran and Reddy \cite{Quarkyonic1} proposed quarkyonic matter—a hybrid state where nucleons and quarks behave as quasi-particles due to a high-density crossover transition in the NS core, which is later extended by Tinaqi and Lattimer \cite{Quarkyonic2} to include beta equilibrium as well. NSs, with their immense gravitational fields, can efficiently capture and retain dark matter (DM). While the true nature of DM remains unknown, numerous theoretical models attempt to explain its observed effects. In this work, we consider neutralino DM—a fermionic particle that interacts with nucleonic matter through the Higgs portal mechanism, providing a potential probe for DM signatures in dense stellar environments \cite{DM_effects1,DM_effects2}.
\section{Formalism}
The equilibrium state of NS is described by the Tolman-Oppenheimer-Volkov (TOV) equation \cite{TOV1, TOV2}. These equations determine the mass ($M$), radius ($R$) of the star. Inside the star, the perturbed fluid and metric can be reduced to four degrees of freedom. Two of which are associated with fluid perturbation, while the other two correspond to space-time perturbation, which are effectively described by the perturbation functions $H_1$, K, W, X \cite{fmode}. These coupled four perturbation equations, along with TOV equations with physical boundary conditions, determine a single frequency $\omega$ solution. The exterior of the star is devoid of matter, and one is left with only the metric perturbations $H_1$ and $K$. These two first-order differential equations can be combined and written as a single second-order differential equation known as the Zerilli equation, given as \cite{Zerelli_eqtn}.
\begin{equation}
	\left[ {d^2 \over dr_\ast^2} + \omega^2 - V_Z(r) \right] Z = 0 \ .
	\label{zerilli}\end{equation}
Here, $r_\ast$ is  the tortoise coordinate  defined by
\begin{equation}
	{d \over dr_\ast } = \left( 1 -{2M \over r} \right)  {d \over dr} \ ,
	\label{tortoise}\end{equation}	
and the $V_Z$ is the effective potential  given by
	\begin{eqnarray}
		&&V_Z(r) = 2 \left( 1 - {2M\over r} \right) \times \nonumber \\
		&& {n^2(n+1)r^3 +
			3n^2Mr^2 +
			9nM^2r + 9M^3 \over r^3(nr+3M)^2 } \ .
	\end{eqnarray}
The equation admits two linearly independent solutions, corresponding to incoming and outgoing waves. For a generic frequency, the physically valid solution inside the star results in a combination of both outgoing and incoming waves at spatial infinity. However, the quasinormal modes of the system are the discrete frequencies at which only purely outgoing waves are present at spatial infinity. For the numerical solution in the exterior of the star, we follow the phase amplitude approach by Anderson et al.\cite{Anderson_1}. They introduced a new dependent variable $\Psi$ according to
\begin{equation}
	Z = \left( 1 - {2 M \over r} \right)^{-1/2} \Psi \ .
	\label{zpsi}\end{equation}
Then we get a new differential equation 
\begin{equation}
	\left[ {d^2 \over dr^2} + U(r) \right] \Psi = 0 ,
	\label{ode}\end{equation}
where
\begin{equation}
	U(r) = \left( 1 - {2 M \over r} \right)^{-2} \left[
	\omega^2 - V_Z(r) + {2M\over r^3} -  {3M^2
		\over r^4} \right] \ .
		\end{equation}  .
The solution consists of two linearly independent solutions of the form 
\begin{equation}
	\psi^\pm = q^{-1/2} \exp \left[ \pm i \int q dr \right] \ ,
	\label{pamsols}\end{equation} 
	the function $\psi^+$ represents an ingoing wave whereas $\psi^-$ is outgoing wave. This gives a non-linear differential equation of  $q(r)$ given as
\begin{equation}
	{1\over 2q} {d^2q \over dr^2} - {3 \over 4q^2} \left( {dq \over dr}
	\right)^2
	+q^2 - U = 0 \ .
	\label{qequation}\end{equation}			
	
The problem for stellar exterior is now reduced to solving the non-linear equation instead of a simple second-order one. We integrate for the exterior
solution along a straight line in the complex $z$-plane.
This line is parametrized by the real distance $\rho$ (from the surface
of the star) with constant phase angle $\theta$.
Then $dz/d\rho = e^{i\theta}$ and $dq/d\rho$ are the derivatives
of $q$ with respect to the real parameter. Thus we get
\begin{equation}
	{dq \over dr} = {dq \over dz} =  e^{-i\theta}{dq \over d\rho},
\end{equation}
which is easily implemented numerically. The ratio of incoming ($A_{in}$) and outgoing amplitude ($A_{out}$) is given as  \begin{equation}
	{A_{\rm out} \over A_{\rm in} } = {Z_S \left[ \left( 1 - {2 M \over
			R} \right) \left(iq - {1\over 2q} {dq \over dr} \right) -
		{ M \over  R^2} \right] - Z_S^\prime \over  Z_S
		\left[ \left( 1 - {2 M \over R} \right)
		\left(iq + {1\over 2q} {dq \over dr} \right) + { M \over
			R^2} \right] + Z_S^\prime } \ .
	\label{ratio}\end{equation}
The $w$-mode of the star is determined as the singular points of this ratio as a function of $\omega$.  

\section{Results}
\label{results} 
\begin{figure*}
\centering
\includegraphics[width=0.93\columnwidth]{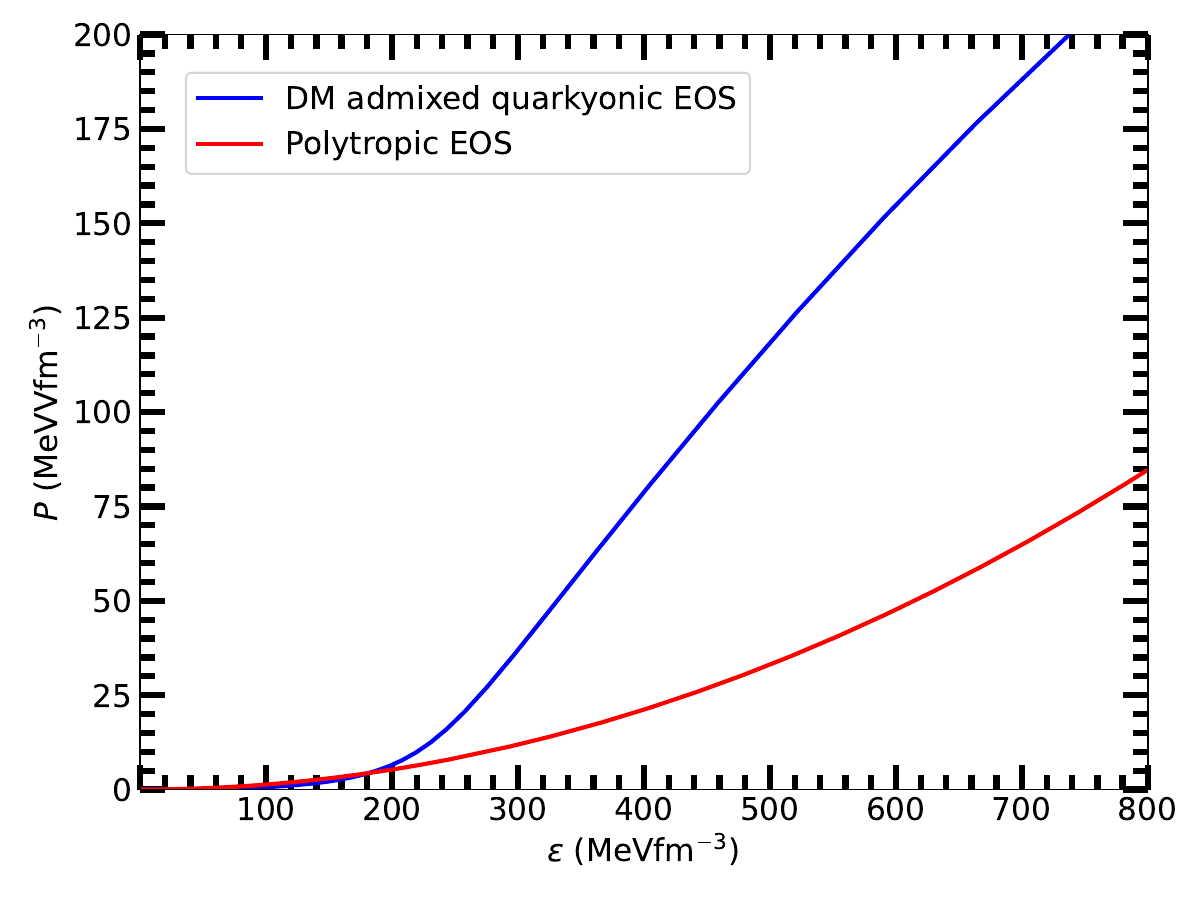}
\includegraphics[width=1.0\columnwidth]{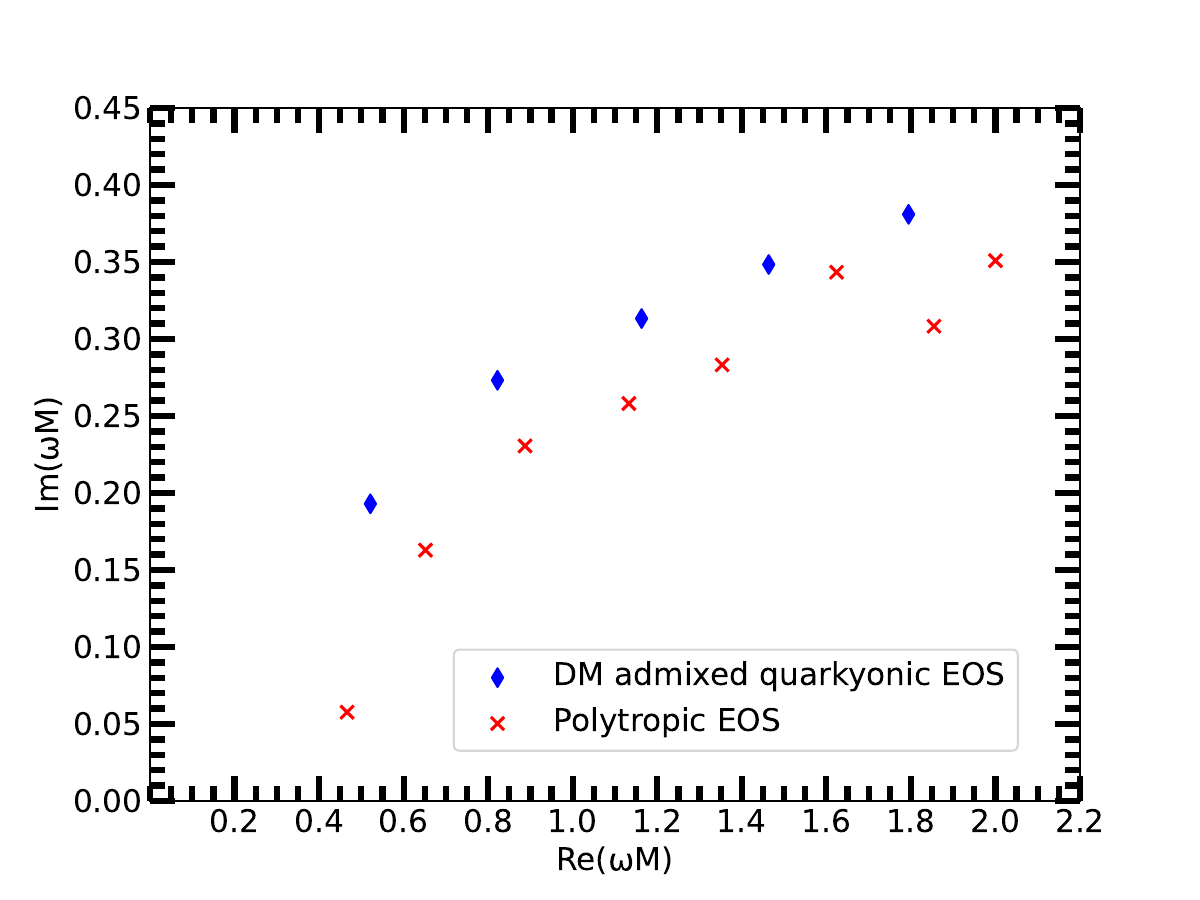}
\caption{(Left Panel) The EOS of DM admixed quarkyonic star for the transition density $n_t = 0.3$ fm$^{-3}$, confinement scale $\Lambda_{cs}= 800$ MeV,  DM fermi momentum $k_f^{DM}$ = 0.03 GeV for IOPB-I parameter set and polytropic star. (Right Panel) The corresponding $w$-mode frequency spectrum for a canonical star in dimensionless form. }
\label{fig1}
\end{figure*}
\begin{table}[h!]
\centering
\caption{Few $w$-mode frequencies (dimensionless) for $l=2$ with polytropic EOS are compared with the DM admixed quarkyonic EOS}
\begin{tabular}{|c|c c|c c|}
\hline
\multirow{2}{*}{Mode} & \multicolumn{2}{c|}{Polytropic} & \multicolumn{2}{c|}{DM-admixed quarkyonic} \\
\cline{2-5}
 & Re($\omega M$) & Im($\omega M$) & Re($\omega M$) & Im($\omega M$) \\
\hline
$w_{1}$ & 0.466165 & 0.057644 & 0.521303 & 0.192982 \\
$w_{2}$ & 0.651629 & 0.162907 & 0.822055 & 0.273183 \\
$w_{3}$ & 0.887218 & 0.230576 & 1.162907 & 0.313283 \\
$w_{4}$ & 1.132832 & 0.258145 & 1.463659 & 0.348371 \\
$w_{5}$ & 1.353383 & 0.283208 & 1.794486 & 0.380952 \\
\hline
\end{tabular}

\label{tab:wmode_merged}
\end{table}

The EOS of a DM admixed quarkyonic star for the case of transition density 
$n_t = 0.3 \,\mathrm{fm}^{-3}$, confinement scale $\Lambda_{cs} = 800 \,\mathrm{MeV}$, and DM Fermi momentum 
$k_f^{DM} = 0.03 \,\mathrm{GeV}$ is shown in the left panel of Fig.~\ref{fig1}. In this framework, the parameters 
$n_t$ and $\Lambda_{cs}$ together regulate the onset and the fraction of quark matter (in our case up ($u$) and down ($d$) quarks) 
appearing in the core of the NS. The model assumes a cross-over type transition between nucleonic and quark 
degrees of freedom, where nucleons, treated as quasiparticles, occupy a finite Fermi shell while the lower momentum states 
are filled by deconfined quarks. This quarkyonic construction leads to a stiffening of the EOS compared to the purely baryonic 
case, since the emergence of quark degrees of freedom enhances pressure at intermediate densities by allowing quarks to drip out of nucleons and occupy low-momentum states. For the baryonic sector of the star, we have employed the IOPB-I parametrization of the relativistic mean field theory \cite{PhysRevC.97.045806}, which treats nucleons (protons and neutrons) as interacting 
through mesonic mean fields ($\sigma$, $\omega$, $\rho$, and $\delta$) with higher-order self- and cross-couplings fitted to reproduce 
empirical nuclear matter and finite nuclei properties throughout the mass table. 

This ensures that the hadronic sector remains consistent with experimental and astrophysical 
constraints at saturation density. On the other hand, the DM contribution to the EOS is entirely governed by its Fermi 
momentum, $k_f^{DM}$, which in the present case is set to $0.03 \,\mathrm{GeV}$, a value motivated by the assumption that the baryonic density inside the NS is about $10^3$ times larger than the average DM density, leading to overall softening of EOS \cite{DDey2}. The right panel of Fig.~\ref{fig1} shows the $w$-mode spectrum for $l=2$ corresponding to the canonical NS
constructed with the EOS considered in our work. The eigen frequencies are obtained as the singularities of 
Eq.~\ref{ratio}, which follows from the condition that only outgoing waves exist at spatial infinity. In order to 
present the results in a dimensionless and compact manner, the frequencies are scaled with the stellar mass. The 
spectrum reveals that both the real and imaginary parts of the frequency are of comparable magnitude, a 
characteristic feature of $w$-modes that distinguishes them from fluid modes such as $p$- or $g$-modes. The real part 
represents the oscillation frequency of the mode, whereas the imaginary part quantifies the damping timescale, 
indicating that $w$-modes are highly damped due to their strong coupling to gravitational radiation.\\
\begin{figure}
\centering
\includegraphics[width=1.0\columnwidth]{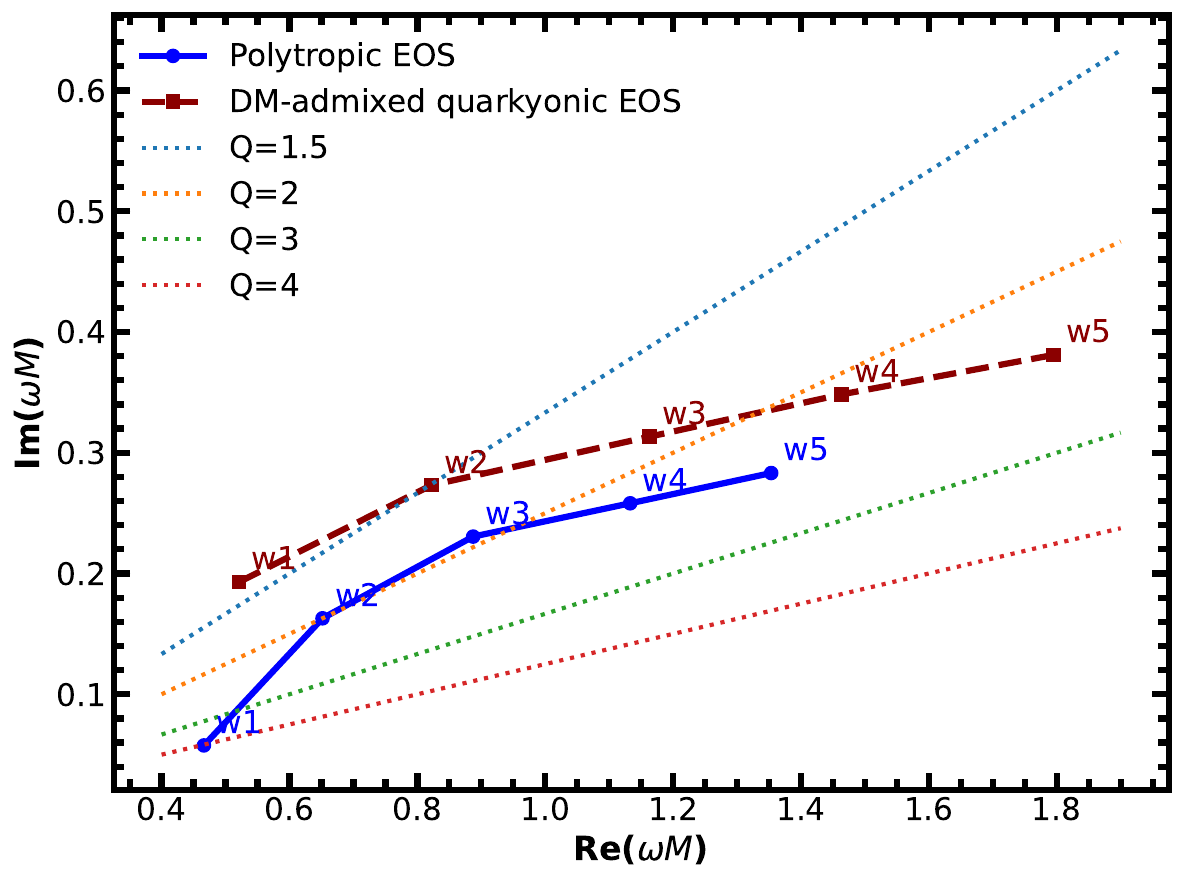}
\caption{(Left Panel) Same as right panel of Figure \ref{fig1}, with the quality factor
$Q = \frac{\mathrm{Re}(\omega)}{2\,|\mathrm{Im}(\omega)|} \,.$
Dotted contours show where $Q = 1.5, 2, 3, 4$.  }
\label{fig2}
\end{figure}

For DM admixed quarkyonic star, a well-defined spectrum emerges, in which the real part of the frequency increases 
almost uniformly with mode number, exhibiting an approximately constant spacing ($\sim$ 0.30) between successive overtones as shown in Table \ref{tab:wmode_merged}. In contrast, the imaginary part of the frequency shows only a mild variation with mode number. For comparison, we have included a simple polytropic EOS defined by $P = K \varepsilon^{2}$ \cite{fmode2}. While this choice of EOS is not physically realistic, it serves to simplify the calculations and provides a useful baseline. As shown in Fig. \ref{fig1}, the DM admixed quarkyonic EOS is significantly stiffer than the polytropic model. This stiffness can be ascribed to the presence of the quarkyonic core and the relativistic magnitude of the speed of sound, which is reached near the transition density $n_t$ where the cross-over between hadronic and quark matter occurs. The admixed DM content, however, has a softening effect, moderating the overall stiffness and resulting in a predicted maximum mass of approximately $2.33M_\odot$ \cite{DDey2}. On the other hand, the parameters of the simple polytropic EOS can be arbitrarily tuned to achieve a wide range of masses and radii; it possesses no intrinsic maximum mass limit. Our model's free parameters, governing the DM core and the quarkyonic transition, are constrained by multi-messenger astrophysical observations  \cite{Quarkyonic1,Quarkyonic2}. The distinct EOS properties are directly reflected in the calculated $w$-mode spectrum, as detailed in Table \ref{tab:wmode_merged} and illustrated in Fig. \ref{fig1}. The spectrum for the realistic DM admixed quarkyonic model exhibits significantly larger imaginary components ($\text{Im}(\omega M)$) compared to the polytropic case. This indicates a stronger coupling to gravitational radiation and thus a faster damping timescale for the oscillations. The real parts of the frequencies ($\text{Re}(\omega M)$), which govern the oscillation period, are also higher for the stiffer EOS but show a less relative change. This suggests that while the fluid perturbations and internal stress-energy distribution significantly affect the spacetime curvature within the star—altering the $w$-mode spectrum—the real part of the frequency is more sensitive to the overall compactness, whereas the imaginary part is highly sensitive to the specific internal structure and EOS stiffness. Since $w$-modes are primarily governed by the curvature of spacetime outside the star. Their spectrum shares profound similarities with the quasi-normal modes of black holes, as both are solutions to the relativistic wave equation in a curved potential well. However, the key difference lies in the presence of a material surface and a complex internal structure in NSs, which modifies the effective potential and introduces this strong dependence on the EOS. The higher damping rates (larger $\text{Im}(\omega)$) for the quarkyonic model indicate that these stars would radiate gravitational wave energy stored in these perturbations more efficiently than their polytropic counterparts.\\

On a different note, from the comparison point of view, we find that the real part of the $w$ mode frequency increases by approximately 12–33\% when moving from the polytropic to the DM-admixed quarkyonic EOS, while the imaginary part (damping) shows an even stronger enhancement, rising by about 35–70\% for higher modes and by more than 230\% for the fundamental mode $w_{1}$. This further suggests that DM admixture systematically shifts the spectrum toward higher frequencies while substantially amplifying damping, especially for the lower-order modes.

The $w$-mode spectrum of the canonical mass star reveals a distinct contrast between the polytropic EOS and the DM-admixed quarkyonic EOS when analyzed through the quality factor 
\[
Q = \frac{\mathrm{Re}(\omega M)}{2\,|\mathrm{Im}(\omega M)|}
\]
and the damping ratio 
\[
R = \frac{|\mathrm{Im}(\omega M)|}{\mathrm{Re}(\omega M)} \, .
\]
As we notice, the polytropic EOS yields higher quality factors ($Q \sim 2{-}4$) and smaller damping ratios ($R \sim 0.12{-}0.26$), corresponding to oscillations which are longer-lived, with multiple cycles before decay, while the DM-admixed quarkyonic EOS produces lower quality factors ($Q \sim 1.3{-}2.3$) and larger damping ratios ($R \sim 0.21{-}0.37$), indicating modes that are heavily damped and short-lived. The Fig. \ref{fig2} with constant-$Q$ contours, clearly shows this separation where the polytropic modes cluster along higher-$Q$ bands. In contrast, DM-admixed modes lie closer to low-$Q$ regions, explaining why higher overtones ($w_6$, $w_7$) survive in the polytropic case but are suppressed in the DM-admixed case. Collectively, these results show that $Q$ and $R$ provide compact and observationally relevant indicators of the internal composition of compact objects, with low-$Q$, high-$R$ $w$ modes serving as potential signatures of exotic components such as dark matter. Similar definitions of the quality factor and its relation to damping have been widely analysed in the quasi-normal mode literature and gravitational-wave ringdown studies \cite{Kokkotas1999,Berti2006}.

\section{Conclusions}
The study of $w$-modes provides an important probe into the space-time dynamics of NSs and complements the information obtained from fluid oscillations. Although these modes are rapidly damped, they are expected to leave a characteristic gravitational-wave signature in the immediate aftermath of stellar collapse, appearing as a short-lived but distinct ringing signal. Since the axial perturbations governing $w$-modes depend on the internal density and pressure distributions, their spectrum is directly linked to the neutron-star EOS.  In particular, the stiffness of the EOS, which can be modified by the presence of exotic phases such as quarkyonic matter or by the admixture of fermionic DM, 
significantly affects both the oscillation frequencies and damping times. Quark matter tends to stiffen the EOS, supporting higher masses and producing larger $w$-mode frequencies, whereas DM generally softens the EOS, reducing stellar compactness and shifting the spectrum accordingly. Despite their fast decay compared to fluid modes, the nearly uniform spacing of the real parts of successive overtones and the slow variation of the imaginary components establish a robust spectral pattern. In addition to that, the quality factor Q and damping ratio R provide clear discriminants between polytropic and DM-admixed quarkyonic stars, with the latter showing systematically lower Q and higher R. These suggest that the observational extraction of $w$ mode lifetimes from gravitational-wave data could offer direct evidence for or against the presence of dark matter in neutron star interiors. Thus, any future detection of $w$-modes through gravitational waves would not only confirm their existence but also serve as a sensitive probe of the EOS at supra-nuclear densities, helping to discriminate between purely hadronic stars and those containing quarkyonic cores or DM admixtures. In this way, $w$-modes could play a decisive role in connecting gravitational-wave astronomy with the fundamental physics of dense matter. A more detailed formalism and further investigations will be addressed in a forthcoming manuscript currently under preparation \cite{Ddey3}. 
\bibliography{wmode_reference}
\bibliographystyle{elsarticle-num.bst}

\end{document}